# Enhancement of Zero-Field Skyrmion Density in [Pt/Co/Fe/Ir]$_2$ Multilayers by FORC

Mangyuan Ma,[1,#] Calvin Ching Ian Ang,[2,#] Yong Li,[1] Weiliang Gan,[2] Wen Siang Lew,[2,*] and Fusheng Ma[1,*]

[1]*Jiangsu Key Laboratory of Opto-Electronic Technology, Center for Quantum Transport and Thermal Energy Science, School of Physics and Technology, Nanjing Normal University, Nanjing 210046, China*
[2]*School of Physical and Mathematical Sciences, Nanyang Technological University, 21 Nanyang Link 637371, Singapore*

[#]These authors contributed equally to this work
*Correspondence to Fusheng Ma, phymafs@njnu.edu.cn
*Joint correspondence to Wen Siang Lew, wensiang@ntu.edu.sg

## Abstract

Magnetic skyrmions are novel topological spin textures on the nanoscale, and significant efforts have been taken to improve their zero-field density at room temperature (RT). In this work, we reported an approach of improving zero-field skyrmion density in [Pt/Co/Fe/Ir]$_2$ multilayers at RT by using the first-order reversal curve (FORC) technique. Firstly, we investigated the nucleation and annihilation mechanism of magnetic skyrmions using polar magneto-optical Kerr effect measurement. Secondly, the FORC technique was used to obtain information on the irreversible or reversible behaviors in the magnetization switching process. It was found from FORC diagram that the magnetization reversal mechanism can be characterized into three stages: (1) reversible labyrinth stripe domains expanding or shrinking stage; (2) irreversible stripe domains fracturing stage; and (3) irreversible skyrmion annihilation stage. At the end, we demonstrated that the zero-field skyrmion density can be highly improved by choosing reversal field from the irreversible stages. Our results have established the FORC measurement as a valuable tool for investigating magnetic multilayers of high skyrmion densities.

# 1 Introduction

Magnetic skyrmions are topologically protected quasi-particles with non-trivial spin texture which could serve as information carriers in novel spintronic devices, such as magnetic racetrack memory, logic gates, spin-torque oscillators, and artificial synapses/neurons [1–5]. In recent years, the existence of skyrmions in magnet/heavy-metal multilayers with perpendicular magnetic anisotropy at room temperature (RT) have been realized resulting from the competition between Dzyaloshinskii-Moriya interaction, dipolar interaction and exchange interaction [6–10]. However, it is still challenging to approach high skyrmion densities without the assistance of external magnetic field at RT. To overcome this obstacle, lots of works have been done [11–17], for instance, generating high density skyrmions from the labyrinth domains at zero-field in Pt/Co/Ta multilayers using an electromagnetic coordinated method [14] and a scanning local magnetic field [16].

The first-order reversal curve (FORC) technique has been proven to be an effective method in probing magnetization reversal mechanism [18–24]. This approach collects a larger data set compared with the complete magnetization hysteresis (*M-H*) loop, from which information about the distribution of switching fields and interaction fields for all of the domains that contribute to the *M-H* loop and the irreversibility of the magnetization switching can be derived, such as the magnetization switching process during a complete reversal in Co/Pt multilayers [25]. We can also enhance the zero-field skyrmion density via analyzing the process of magnetization reversal and find an irreversible stage where the higher density of skyrmions may observed at the corresponding remanent states, so, a thorough understanding of the magnetization switching processes involved in skyrmion-hosting materials must first be obtained.

In this work, we describe the magnetization reversal process during an out-of-plane magnetic field sweep from zero field to negative saturation field as three distinct stages. Firstly, a predominantly reversible behavior is observed due to the labyrinth stripe domains expanding or shrinking in width without changing their topology. Secondly, as the applied field increase, the stripe domains fracture into multiple skyrmions. Finally, an irreversible reversal process corresponding to the annihilation of skyrmions. More importantly, we present an effective approach to enhance the zero-field skyrmions density at RT by analyzing the characteristics of the FORC diagram, which is of great significance for the successful application of skyrmions for practical applications.

## 2 Experimental method

The multilayer film stack of Ta(5)/Ir(2)/[Pt(1)/Co(0.5)/Fe(0.4)/Ir(0.9)]$_2$/Ta(2) (the number in parentheses are the nominal layer thickness in nanometer) were deposited on a thermally oxidized silicon wafer using direct current magnetron sputtering techniques at RT. A base pressure of $1 \times 10^{-7}$ Torr or better is reached before sputtering is initiated. Argon gas was used during the sputtering process at 2.0 mTorr for all materials except Co which was sputtered at 3.0 mTorr. The deposition rates for Ta, Pt, Co, Fe, and Ir were 0.62, 0.68, 0.26, 0.30, and 0.37 Å/s, respectively. The bottom Ta layer was used as an amorphous and adhesive underlayer. The additional Ir underlayer was added to provide a similar interface for the repeating layers. Additional capping layer of Ta was used as a protective layer against oxidation.

The out-of-plane *M-H* loop and FORCs of the [Pt/Co/Fe/Ir]$_2$ multilayers were measured by the MagVision Kerr microscopy system in the polar mode with the magnetization imaged simultaneously. To remove the non-magnetic background signal, differential imaging was performed. Due to the longer measurement durations needed for FORC, piezo motors were used to compensate for the sample drift.

## 3 Results

A full *M-H* loop under an out-of-plane magnetic field $H_z$ was firstly measured using the polar magneto-optical Kerr effect (MOKE) at RT as shown in Fig. 1. The red and blue solid lines represent the ascending and descending branches, respectively. The insets show the MOKE images which correspond to the out-of-plane magnetization configurations at selected fields. The whole magnetization reversal process can be understood from the variation of the magnetic domain configurations. As the ascending and descending branches are reversible, we will explain the variation of magnetic domains along the ascending branch. We divide the process into the polarized (from -20 to 0 Oe) and un-polarized (from 0 to 20 Oe) parts. In the polarized part, starting from the negatively saturated state as shown in image 1, the magnetic domain configurations gradually transfer to complex domains of long labyrinth stripes. Firstly, certain regions of the [Pt/Co/Fe/Ir]$_2$ multilayers usually containing defects which lower the energy barrier for magnetization reversal. These regions will be reversed in the form of circular domains, i.e. the presence of magnetic skyrmions (image 2), as a result of the competition between DMI and demagnetizations [28–30]. Secondly, these skyrmion-like domains expand in a non-symmetrical manner and become into short labyrinth stripes with increasing $H_z$. In the meantime, there are also the assemble presence of short stripes, see image 3. After that, the short stripes lengthen continuously in random directions (image 4) and reach a up/down magnetization balanced remanent configuration at zero field, see image 5. In the un-polarized part, the magnetic domains change roughly in an inverse process of the aforementioned process. The long stripes firstly shorten with the field increases from zero and approach a complex configuration comprised of both stripes and skyrmions, see image 6. Then, the stripes continuously shrink and also break into skyrmions simultaneously (image 7). Image 9 shows a pure skyrmion configuration with all the stripes disappear. Finally, the skyrmions gradually annihilated and a positively saturated state arrives when the field is larger than 15 Oe. To be note that there is hardly any skyrmions in the magnetization configuration at zero-field as shown in image 5 of Fig. 1.

To improve the skyrmion density at zero-field, the FORC technique was adopted to investigate the magnetization reversal characteristics of the [Pt/Co/Fe/Ir]$_2$ multilayers. In particular, the details of the reversible and irreversible behaviors of the magnetization switching process with the history of the field $H_z$. The FORC measurement can be carried out along both the ascending branch and the descending branch of the *M-H* loop [31], respectively. In view of the symmetrical reversal processes along the two branches, in this work, we did the FORC measurement in the manner as defined in Fig. 2a. Firstly, a positive field of $H_{sat}$, large enough to saturate the sample was applied, and then, $H_z$ decreases to the selected fields of $H_R$ (red dots in Fig. 2(a)), from which the field increases to $H_{sat}$. These minor $H_{sat}$ - $H_R$ - $H_{sat}$ *M-H* loops form a whole FORCs together. Figure 2(b) shows the measured FORCs with the reversal field $H_R$ selected in a constant interval of 3 Oe, in which the outer boundary delineates the major *M-H* loop [32–34]. The insets show the zoomed - in views of two portions of the FORCs, which indicate the reversible and irreversible regions of the magnetization. To better understand the minor $H_{sat}$ - $H_R$ - $H_{sat}$ *M-H* loops, the FORCs are usually analyzed by performing Fourier transform to a FORC diagram [32, 35]. The magnetization can be traced out from the FORCs of Fig. 2(b) as a two-variable function $M(H_R, H_z)$ in the $(H_R, H_z)$ plane as shown in Fig. 2(c). The FORC distribution function, a FORC diagram, can be calculated by [19, 33, 36, 37]:

$$\rho = -\frac{1}{2}\frac{\partial^2 M(H_R, H_z)}{\partial H_R \, \partial H_z} \mathrm{d}H_z \tag{1}$$

Figure 2(d) is the contour map plotting of $\rho$ [31].

The FORC diagram reveals the occurrence of reversible and irreversible changes in the reversal process: $\rho = 0$ for the pure reversible changes; while $\rho \neq 0$ for the irreversible changes [25, 31, 38]. To understand the reversible and irreversible switching process, we mainly focused on the third quadrant of the FORCs diagram since the selected reversal fields $H_R < 0$. According to the values of $\rho$ in Fig. 3(a), we divide the FORC diagram into three regions as marked by A, B and C, respectively. Region A (0 Oe ≤ -$H_R$ ≤ 7 Oe) is a nearly featureless region with the value of $\rho$ close to zero. Here, the almost zero value of $\rho$ implies a reversible changes and the corresponding parts of the FORCs are mostly overlapped as the top inset in Fig. 2(b). The magnetization reversal processes are reversible with the width of the stripe domains shrinking with field $H_z$ and the number of domains keeping unchanged. In contrast, in region B (7 Oe ≤ -$H_R$ ≤ 11 Oe), the value of $\rho$ changes highly with the presence of a maximum at (-$H_R$, -$H_z$) = (9.6Oe, 9.6Oe). The field-dependent variation of $\rho$ indicates the existence of an irreversible behavior in the magnetization switching, which corresponds to a non-overlapped part of the FORCs as the bottom inset in Fig. 2(b). The irreversible process corresponds to the fission of stripe domains to skyrmions. Lastly, in region C (11 Oe ≤ $-H_R$ ≤ 15 Oe), the $H_R$ closing to the negative saturation field, a variation of $\rho$ was observed with the presence of a negative valley and positive peak at $H$ = -10.0 and -8.5 Oe, respectively. The pair of negative-positive peaks indicates the annihilation of skyrmions.

The dependence of skyrmion densities at $H_z = H_R$ and $H_z = 0$ on $H_R$ is plotted in Fig. 3(e). It is found that the zero-field skyrmion density firstly increases when the $H_R$ increases from region A to B, and approaches a maximum value of 0.05 μm$^{-2}$ where the skyrmion densities at the corresponded $H_R$ is also the largest. Then, the zero-field skyrmion density decreases with the $H_R$ increasing from region B to C. To further correlate the degree of irreversibility quantified by $\rho$ from the FORC diagram with the enhancement of zero-field skyrmion density at remanent, we calculated the maximum $\rho$ value $\rho_{max}$ for each $H_R$ as open squares in Fig. 3(e). The $H_R$-dependent variation of $\rho_{max}$ exhibits a similar trend to that of skyrmion densities, i.e. firstly increasing to a maximum value in region B and then decreasing from there when the $H_R$ increases from region B to C. Furthermore, the maximum value of the $\rho_{max}$ and the zero-field skyrmion density appears at a very similar $H_R$. Therefore, the improvement of zero-field skyrmion density can be directly correlated to the irreversibility $\rho$ from the FORC diagram. To maximize the zero-field skyrmion density, we can choose a reversal field $H_R$ corresponding to the maximum value of $\rho$ in the $H_{sat}$ - $H_R$ - 0 process.

## 4 Conclusion

In summary, our MOKE and FORC measurements reveal three distinct magnetization switching mechanisms in our sample of [Pt/Co/Fe/Ir]$_2$ multilayers: (1) stripe domains expanding and shrinking (reversible); (2) stripe domains fission (irreversible); and (3) skyrmions annihilation (reversible). We have found an effective method to highly enhance the zero-field skyrmion density using the FORC technique:

by analyzing the features of the FORC diagram, a region of reversal field is found, in which the corresponding value of $\rho$ is maximum, and the density of skyrmion is highest, then reduce the applied field from the reversal field to zero, which is the state with the largest zero-field skyrmion density at RT. Our results also demonstrate that the FORC technique is a sensitive method to probe the detailed information during the magnetization switching process in perpendicular magnetic multilayers.

**Acknowledgements**

We thank Anthony Tan for his help on understanding the FORC diagrams. Work performed at Nanjing Normal University was supported by the National Natural Science Foundation of China (Grant No. 11704191), the Natural Science Foundation of Jiangsu Province of China (Grant No. BK20171026), the Jiangsu Specially-Appointed Professor, and the Six-Talent Peaks Project in Jiangsu Province of China (Grant No. XYDXX-038). Work performed at Nanyang Technological University was supported by a NRF-CRP Grant (CRP9-2011- 01), a RIE2020 ASTAR AME IAF-ICP Grant (No.I1801E0030) and an ASTAR AME Programmatic Grant (No. A1687b0033). W.S.L is a member of the SG-SPIN Consortium.

long-term thermal aging on magnetization process in low-alloy pressure vessel steels using first-order-reversal-curves. AIP Adv 7:056002-1-056002–6. https://doi.org/10.1063/1.4973605

35. Gräfe J, Schmidt M, Audehm P, et al (2014) Application of magneto-optical Kerr effect to first-order reversal curve measurements. Rev Sci Instrum 85:. https://doi.org/10.1063/1.4865135
36. Dobrotă CI, Stancu A (2013) What does a first-order reversal curve diagram really mean? A study case: Array of ferromagnetic nanowires. J Appl Phys 113:043928-1-043928–11. https://doi.org/10.1063/1.4789613
37. Martínez-García JC, Rivas M, Lago-Cachón D, García JA (2014) First-order reversal curves analysis in nanocrystalline ribbons. J Phys D Appl Phys 47:1–7. https://doi.org/10.1088/0022-3727/47/1/015001
38. Roberts AP, Liu Q, Rowan CJ, et al (2006) Characterization of hematite (α-Fe2O3), goethite (α-FeOOH), greigite (Fe3S4), and pyrrhotite (Fe7S8) using first-order reversal curve diagrams. J Geophys Res Solid Earth 111:1–16. https://doi.org/10.1029/2006JB004715
10

**Figures**

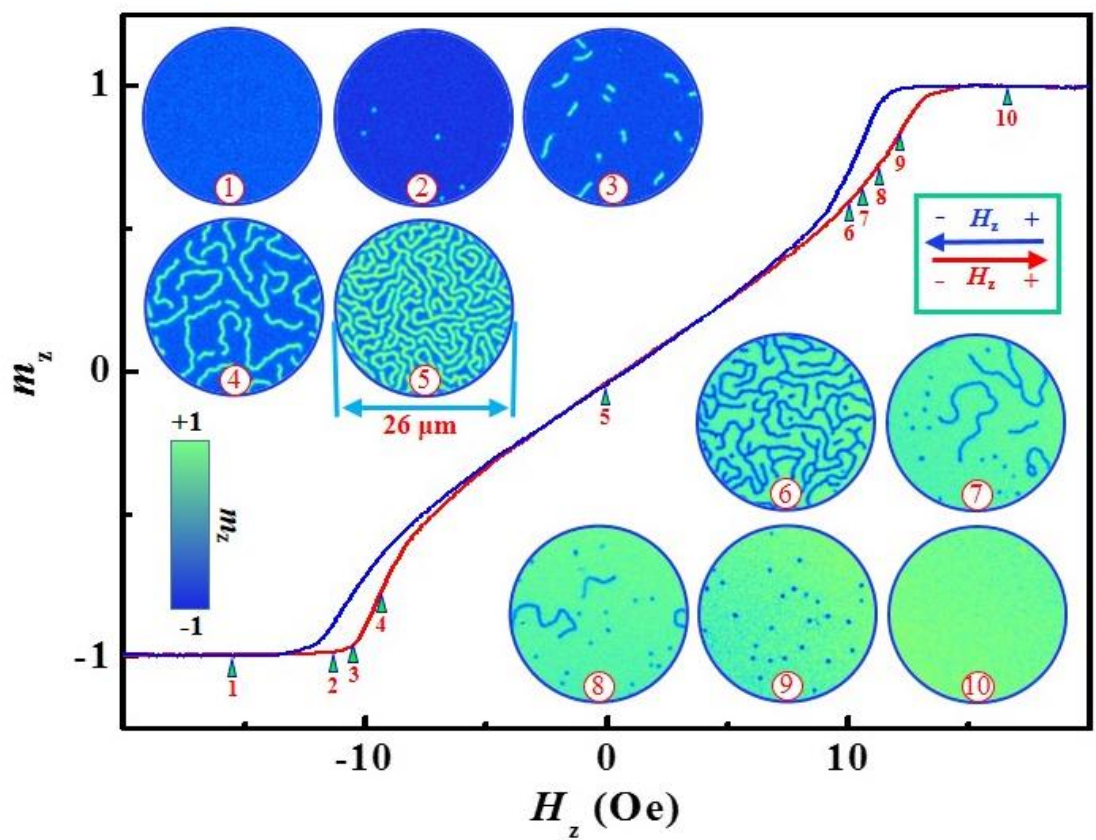

**Figure 1** Evolution of the out-of-plane magnetization configurations in [Pt/Co/Fe/Ir]$_2$ multilayers at RT. The red and blue solid lines represent the ascending and descending branches of the *M-H* loop, respectively. Insets are differential polar-MOKE images showing the field evolution of magnetic domains at selected fields as labelled by triangles along the ascending branch of the *M-H* loop. A circular region with a diameter of 26 μm was prepared from the measured Kerr images.

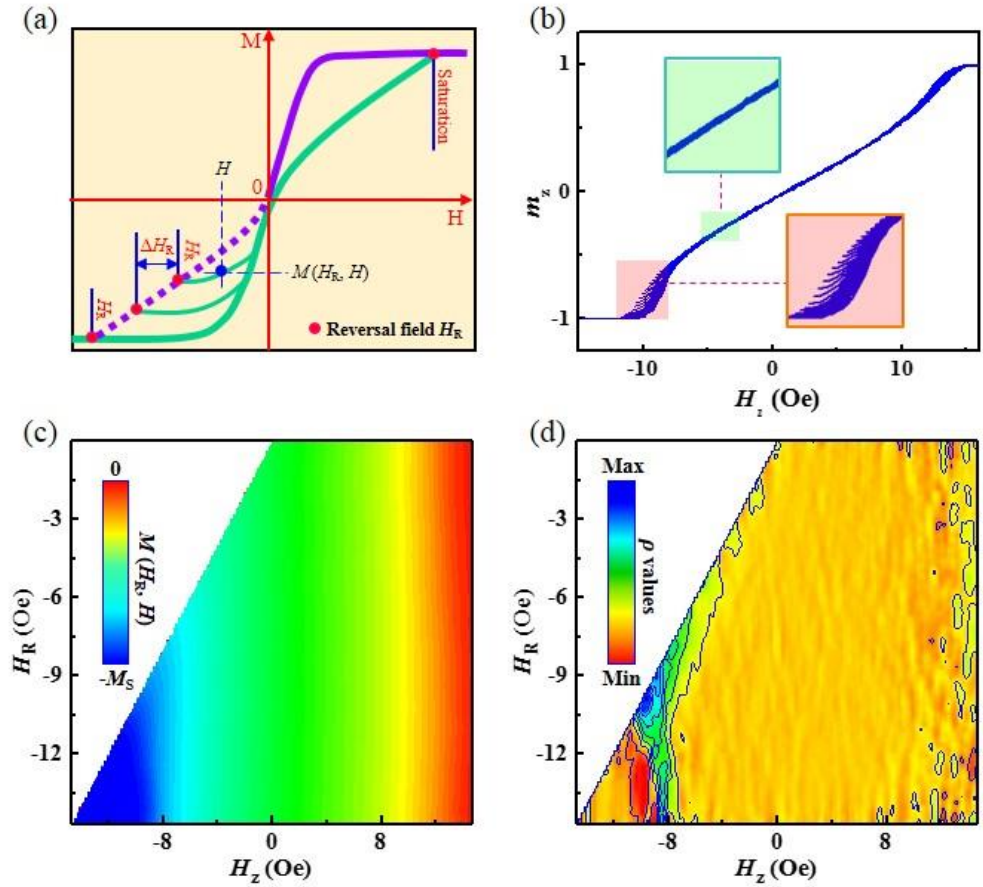

**Figure 2** FORC measurement and the magnetization irreversibility. (a) Schematic of FORC measurement: the red dots represent the magnetizations at the reversal fields $H_R$, the green solid lines represent the field sweeping paths, and $\Delta H_R$ represents an interval between reversal fields of 3 Oe. (b) Polar-MOKE measured FORCs of [Pt/Co/Fe/Ir]$_2$ multilayers. Insets are zoomed-in views of different portions of FORCs. (c) 2D map of magnetization values obtained by measuring the magnetization along each FORC. (d) FORC diagram, a contour plot of the corresponding FORC distribution $\rho$.

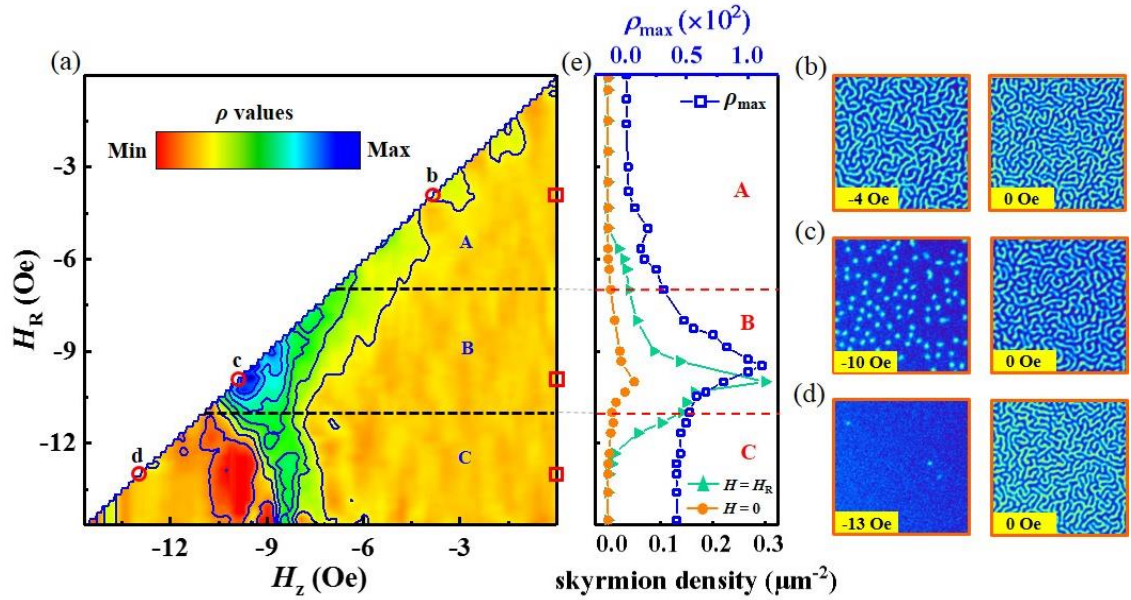

**Figure 3** Magnetization irreversibility and domain topology modification. (a) FORC diagram divided in to three regions according to the degree of the reversible or irreversible behaviors. (b)-(d) Polar-MOKE images (the side length is 20 μm) of the magnetic domain structures measured at selected reversal fields $H_R$ = -4, -10, -13 Oe corresponding to aforementioned three regions (left column, open circles in (a)), and subsequently measured at the corresponded remanent states (right column, open squares in (a)). (e) Skyrmion density measured at selected reversal fields $H_R$ (green solid line with symbols of triangles) and subsequently measured at the remanent states when field changed from $H_R$ to 0 (origin solid line with symbols of dots), respectively. The maximum values of the FORC distribution $\rho$ measured at different reversal fields represented by the blue solid line with symbols of open squares.